\documentclass{aa}
\usepackage{graphicx}
\usepackage{txfonts}
\begin{document}
   \newpage
   \title{A model of accelerating dark energy in decelerating gravity}

   \author{M. Roos}

   \institute{Departments of Physical Sciences and
          Astronomy, FIN-00014 UNIVERSITY OF HELSINKI\\
              \email{matts dot roos at Helsinki dot Fi}}


  \abstract
   {The expansion of the Universe is
accelerated as testified by observations of supernovae of type Ia at
varying redshifts. Explanations of this acceleration are of two
kinds: modifications of Einstein gravity or new forms of energy,
coined dark energy. An example of modified gravity is the braneworld
Dvali-Gabadadze-Porrati (DGP) model, an example of dark energy is
Chaplygin gas. Both are characterized by a cross-over length scale
$r_c$ which marks the transition between physics occurring on our
four-dimensional brane, and in a five-dimensional bulk space.}
   {Assuming that the scales $r_c$ in the two models are the same,
  we study Chaplygin gas dark energy in both self-accelerating and
  self-decelerating flat DGP geometries. The self-accelerating
  branch does not give a viable model, it causes too much
  acceleration.}
  {We derive the Hubble
  function and the luminosity distance for the self-decelerating
  branch, and then fit a compilation of 192 SNeIa magnitudes and
  redshifts. This determines a confidence region in the space of the
  three parameters of the model.}
   {Our model with the self-decelerating branch in flat space fits
   the supernova data as successfully as does the $\Lambda CDM$
   model, and with only one additional parameter.}
   {In contrast to
   the $\Lambda CDM$ model, this model needs no fine-tuning, and it can explain the
   coincidence problem. It is unique in the sense that it cannot be
   reduced to a cosmological constant model in any other limit of
   the parameter space than in the distant future. If later tests
   with other cosmological data are successful, we have here a first
   indication that we live in a five-dimensional braneworld.}


   \maketitle

\section{Introduction}

The demonstration by SNeIa that the Universe is undergoing an
accelerated expansion has stimulated a vigorous search of models to
explain this unexpected fact. Since the dynamics of the Universe is
conventionally described by the Friedmann equations which follow
from the Einstein equation in four dimensions, all modifications ultimately affect
the Einstein equation.

The left-hand-side of the Einstein equation encodes the geo\-metry of
the Universe in the Einstein tensor $G_{\mu\nu}$, the
right-hand-side encodes the energy content in the stress-energy
tensor $T_{\mu\nu}$. Thus modifications to $G_{\mu\nu}$ imply some
alternative geo\-metry, modifications in $T_{\mu\nu}$ involve new
forms of energy densities that have not been observed, and which
therefore are called dark energy.

A well-studied model of modified gravity is the
Dvali-Gabadadze-Porrati (DGP) braneworld model (Dvali \cite{Dvali}, Deffayet \cite{Deffayet}) in which our four-dimensional world is a FRW brane embedded in a five-dimensional Minkowski bulk.
The model is characterized by a cross-over length scale  $r_c$
such that gravity is a four-dimensional theory at scales $a\ll r_c$
where matter behaves as pressureless dust. In the self-accelerating DGP branch gravity "leaks out"
into the bulk at scales $a \gg r_c$ and the cosmology approaches the
behavior of a cosmological constant. To explain the accelerated
expansion which is of recent date ($z\approx 0.5$ or $a\approx 2/3$), $r_c$ must be of the order of 1.
In the self-decelerating DGP branch gravity "leaks in" from the bulk at scales $a\gg r_c$, counteracting the observed dark energy acceleration.

Another well-studied model introduces into $T_{\mu\nu}$ the density $\rho_{\varphi}$ and pressure $p_{\varphi}$ of a fluid called Chaplygin gas
(Kamenshchik \cite{Kamenshchik}, Bilic \cite{Bilic}) following historical work in aerodynamics (Chaplygin \cite{Chaplygin}). This model is
similar to the DGP model in the sense that it is also characterized
by a cross-over length scale below which the gas behaves as
pressureless dust, and above which it approaches the behavior of a
cosmological constant. This length scale is expected to be of the
same order of magnitude as the $r_c$ scale in the DGP model.

Both the self-accelerating DGP model in flat space and the standard Chaplygin gas model have problems
fitting present supernova data, as demonstrated by Davis (\cite{Davis}). In the standard Chaplygin gas model the Jeans instability of perturbations behaves like CDM fluctuations in the dust-dominated stage ($a\ll r_c$), but disappears in the acceleration stage ($a\gg r_c$). The combined effect of suppression of perturbations and non-zero Jeans length leads to a strong ISW effect and thus of loss of power in CMB anisotropies (Amendola \cite{Amendola}, Bento \cite{Bento}).
This has led to generalizations to higher-dimensional braneworld models which appear less motivated, and which require more parameters.

We, instead, combine the standard DGP model with the standard Chaplygin gas model.

This paper is organized as follows. In Section 2 we discuss how to identify the cross-over scales in the DGP and Chaplygin gas models. This idea is motivated by the
similarities in the asymptotic properties of the models, and was first presented in (Roos \cite{Roos}). In Section 3 we discuss the flat-space self-accelerating basic DGP model with and without standard Chaplygin gas dark energy, and in Section 4 we turn to the self-decelerating DGP model combined with standard Chaplygin gas. In Section 5 we summarize our results, and in Section 6 we discuss them and conclude.

\section{Cross-over scales}

On the four-dimensional brane in the DGP model, the action
of gravity is proportional to $M^2_{Pl}$ whereas in the bulk it is
proportional to the corresponding quantity in 5 dimensions, $M^3_5$.
The cross-over length scale is defined as
\begin{equation}
r_c=\frac{M^2_{Pl}}{2M^3_5}\ .\label{rc}
\end{equation}
It is customary to associate a density parameter with this,
\begin{equation}
\Omega_{r_c}=(2r_c H_0)^{-2}.\label{Omrc}
\end{equation}

The Friedmann equation in the DGP model may be written (Deffayet \cite{Deffayet})
\begin{equation}
H^2-\frac k{a^2}-\epsilon\frac 1 r_c\sqrt{H^2-\frac
k{a^2}}=\kappa\rho~,\label{Friedm}
\end{equation}
where $a=(1+z)^{-1},\ \kappa=8\pi G/3$, and $\rho$ is the total cosmic fluid energy density $\rho=\rho_m+\rho_{\varphi}$. Clearly the standard FRW cosmology is recovered in the limit $r_c\rightarrow\infty$ and $\rho_{\varphi}\rightarrow 0$. In the following we shall only consider $k=0$ flat geometry. The \emph{self-accelerating branch} corresponds to $\epsilon=+1$, the \emph{self-decelerating branch} to $\epsilon=-1$.

Since ordinary matter does not interact with Chaplygin gas, one has separate continuity equations for the energy densities $\rho_m$ and $\rho_{\varphi}$, respectively. In DGP geometry the continuity equations have the same form as in FRW geometry (Deffayet \cite{Deffayet}),
\begin{equation}
\dot\rho+3H(\rho+p)=0\ .
\end{equation}
Pressureless dust with $p=0$ then evolves as $\rho_m(a)\propto a^{-3}$.

The Chaplygin gas pressure is $p_{\varphi}=-A/\rho_{\varphi}$, where $A$ is a constant with the dimensions of energy density squared. The continuity equation for Chaplygin gas is then
\begin{equation}
\dot\rho_{\varphi}+3H\left(\rho_{\varphi}-\frac{A}{\rho_{\varphi}} \right)=0~,
\end{equation}
which integrates to
\begin{equation}
\rho_{\varphi}(a)=\sqrt{A+B/a^6}~,\label{rhoch}
\end{equation}
where $B$ is an integration constant. Thus this model has two free
parameters. Obviously its limiting behavior is
\begin{equation}
\rho_{\varphi}(a)\propto\frac{\sqrt{B}}{a^{3}}~~ {\rm for}~~ a~\ll \left(\frac B A\right)^{1/6},~\rho_{\varphi}(a)\propto -p~~  {\rm for}~~ a\gg \left(\frac B A\right)^{1/6}.
\end{equation}

Our \emph{first central assumption} is that the DGP model cross-over scale $r_c$ and the Chaplygin gas cross-over scale $(B/A)^{1/6}$ are about the same. If we choose the proportionality
\begin{equation}
\left(\frac B A\right)^{1/6}=r_c H_0=2\sqrt{\Omega_{r_c}}\
,\label{BA}
\end{equation}
this permits rewriting Eq.~(\ref{rhoch}) in a dimensionless form.
Making use of Eq.~(\ref{Omrc}),  $\rho_{\varphi}$ becomes
\begin{equation}
\rho_{\varphi}(a)=H_0^2\kappa^{-1}\Omega_A \sqrt{1+(4\Omega_{r_c}a^2)~^{-3}}~,\label{rhophi}
\end{equation}
where we have replaced the energy density $\sqrt{A}$  by the dimensionless density parameter $\Omega_A=H_0^{-2}\kappa\sqrt{A}$.

The identification of the two cross-over scales evidently reduces the number of free parameters in Eq.~(\ref{Friedm}) by one: they are $\Omega_{r_c},~\Omega_A$ and $\Omega_m=\kappa\rho_m H_0^{-2}$.

\section{DGP gravities with and without Chaplygin gas}

Let us now return to the Friedmann equation (\ref{Friedm}) and solve it for the expansion history $H(a)$. Substituting $\Omega_{r_c}$ from Eq.~(\ref{Omrc})~, $\rho_{\varphi}(a)$ from Eq.~(\ref{rhophi}), and $\Omega_m=\Omega_m^0 a^{-3}$, it becomes
\begin{equation}
\frac {H(a)}{H_0}=\epsilon\sqrt{\Omega_{r_c}}+\left[\Omega_{r_c}+\Omega_m^0 a^{-3}
+\Omega_A\sqrt{1+(4\Omega_{r_c}a^2)~^{-3}}\right]^{1/2}.\label{Ha}
\end{equation}
Note that $\Omega_{r_c}$ and $\Omega_A$ do not evolve with $a$. In the limit $a\ll r_c$ this equation reduces to two terms which evolve as $a^{-3}$, thus behaving as dust with density parameter $\Omega_m^0 +\Omega_A/(4\Omega_{r_c})^3$. In the limit $a\gg r_c$, Eq.~(\ref{Ha}) describes a model with a cosmological constant $\Omega_{\Lambda}\equiv -\sqrt{\Omega_{r_c}}+\sqrt{\Omega_{r_c}+\Omega_A}$.

At present, when $a=1$ and $H=H_0$, we solve it for $\Omega_m^0$,
\begin{equation}
\Omega_m^0=1-2\epsilon\sqrt{\Omega_{r_c}}-\Omega_A\sqrt{1+(4\Omega_{r_c}a^2)~^{-3}}.\label{Om0}
\end{equation}
In the well-studied standard self-accelerating DGP model, $\epsilon=+1$ and $\Omega_A=0$, so that
\begin{equation}
\Omega_m^0=1-2\sqrt{\Omega_{r_c}}.\label{DGPflat}
\end{equation}
This equation represents the condition for flatness, and corresponds to the linear relation $\Omega_m^0+\Omega_{\Lambda}=1$ in the $\Lambda CDM$ model. Here, however, Eq.~(\ref{DGPflat}) is nonlinear, causeing $\Omega_m^0$ always to be smaller than in the $\Lambda CDM$ model. This is the reason why the standard self-accelerating DGP model is a worse fit to SNeIa data than the $\Lambda CDM$ model [cf. eg. Davis (\cite{Davis}), Rydbeck (\cite{Rydbeck})]. The failure has led to studies of various generalized DGP models implying higher-dimensional bulk spaces and additional free parameters that detract from its original simplicity and elegance.

The inclusion of the Chaplygin gas term with $\Omega_A>0$ in Eq.~(\ref{Om0}) leads a further reduction in the value of $\Omega_m^0$, and thus to an even worse fit to SNeIa data.
DGP self-acceleration and Chaplygin gas dark energy simply yield too much acceleration, separately as well as in combination. In the next Section we therefore turn to what represents our \emph{second central assumption}, self-decelerating DGP gravity with $\epsilon=-1$. Its expnasion history is still given by Eq.~(\ref{Ha}) and its flat-space condition by Eq.~(\ref{Om0}). The physics then changes, as one can see best in Eq.~(\ref{Om0}), where the two last terms get opposite signs. (The case with $\Omega_A=0$ is not interesting here, because it does not lead to any acceleration.)

\section{Data and method of analysis}
The data we use to test this model are the same 192 SNeIa as in the compilation used by Davis (\cite{Davis}) which is a combination of the "passed" set in Table 9 of Wood-Vasey (\cite{Wood}) and the "Gold" set in Table 6 of Riess (\cite{Riess}).

We are sceptical about using CMB and BAO power spectra, because they have been derived in FRW geometry. SNeIa data are, however, robust for our analysis, since the distance moduli are derived from light curve shapes and fluxes, that do not depend on the choice of cosmological models. In one of our fits we nevertheless include a value for $\Omega_m^0$ as a Gaussian prior which we take from Table 2 of Tegmark (\cite{Tegmark}), who has obtained it in a multi-parameter fit to WMAP and SDSS LRG data. Tegmark's value is $\Omega_m^0=0.239+0.018/-0.017$, but we do not use these $1\sigma$ errors which have been obtained by marginalizing over all other parameters, and which would constrain our fit too strongly. We take the $\Omega_m$ prior to have a large error, $\Delta\Omega_m^0=0.09$, in order not to bias the conclusions from the SNeIa data set. We do not marginalize, but quote full three-dimensional confidence regions: a $1\sigma$ error then corresponds to a contour at $\chi^2_{best} + 3.54$ around the best value $\chi^2_{best}$.

The Davis' compilation lists magnitudes $\mu_i$, magnitude errors $\Delta\mu_i$ for SNeIa at redshifts $z_i,\ i=1,192$. We compute model magnitudes
\begin{equation}
\mu(z_i,\Omega_m,\Omega_{r_c},\Omega_A)= 5~ \rm{Log}[d_L(z_i,\Omega_m,\Omega_{r_c},\Omega_A)]+25\ ,
\end{equation}
where the luminosity distance in Mpc at redshift $z_i$ is
\begin{equation}
d_L(z_i,\Omega_m,\Omega_{r_c},\Omega_A)=(1+z_i)\int_0^{z_i}\frac{dz}{H(z)}\ ,
\end{equation}
where $H(z)$ is given by Eq.~(\ref{Ha}) with $\epsilon=-1$.

We then search in the parameter space for a minimum of the $\chi^2$ sum
\begin{equation}
\chi^2=\sum_{i=1}^{192} \left(\frac{\mu_i-\mu(z_i,\Omega_m,\Omega_{r_c},\Omega_A)}{\Delta\mu_i}\right)^2 + \left(\frac{0.24-\Omega_m}{\Delta\Omega_m}\right)^2.\label{chi}
\end{equation}
Occasionally we do not include the last term.

The calculations are done with the classical CERN program MINUIT (James and Roos \cite{James})
which delivers $\chi^2_{best}$, parameter errors, error contours and parameter correlations.

\section{Results}

Our first fit to determine the region in $\Omega_m^0,\Omega_{r_c},\Omega_A$-space, required by the SNeIa data, keeps all the parameters free, except that $\Omega_m^0$ is restricted to positive values and an upper limit is imposed on $\Omega_A$. We do not include the last term in Eq.~(\ref{chi}). We find a solution at
\begin{equation}
\Omega_m^0=0.40^{+0.13}_{at~lower~limit},~\Omega_{r_c}=1.22^{+0.26}_{-0.71},\
\Omega_A=3.7^{at~upper~limit}_{-1.0}
\end{equation}
with $\chi^2=195.1$. We have checked that the standard $\Lambda CDM$ model fits the same data with the (insignificantly) higher value $\chi^2=195.6$ (as was also found by Davis \cite{Davis}).

Since the confidence region exceeds the limits fixed for $\Omega_m^0$  and $\Omega_A$, obviously $\chi^2$ is very insensitive to values of these parameters near their limits. Moreover, the $\vert (\Omega_m^0,\Omega_{r_c})\vert $ correlation coefficient is $0.97$, thus one of those parameters is almost superfluous. In that sense, our model appears almost as a two-parameter model. One could actually fix $\Omega_A$ at an arbitrary value, but that would be an assumption \emph{ad hoc}.

To cure these fitting problems, we proceed instead to include the last term in Eq.~(\ref{chi}) as a weak prior, choosing $\Delta\Omega_m$ as large as possible, in this case $\Delta\Omega_m^0=0.09$ . This has the effect of neatly putting the error contours well inside all imposed (and now unnecessary) limits, and reducing the correlation coefficient, $\vert (\Omega_m^0,\Omega_{r_c})\vert $ to $0.87$. We now find as best solution the parameter values
\begin{equation}
\Omega_m^0=0.26\pm 0.16,~\Omega_{r_c}=0.82^{+0.69}_{-0.22},~\Omega_A=2.21^{+0.50}_{-0.22}\label{res}
\end{equation}
with $\chi^2=195.5$.

In Fig.1 we plot the best fit confidence region in the $(\Omega_m^0,\Omega_{r_c})$-plane, a banana-shaped closed contour . A cross in the Figure marks the point of best fit.

\begin{figure*}
   \includegraphics[width=9cm]{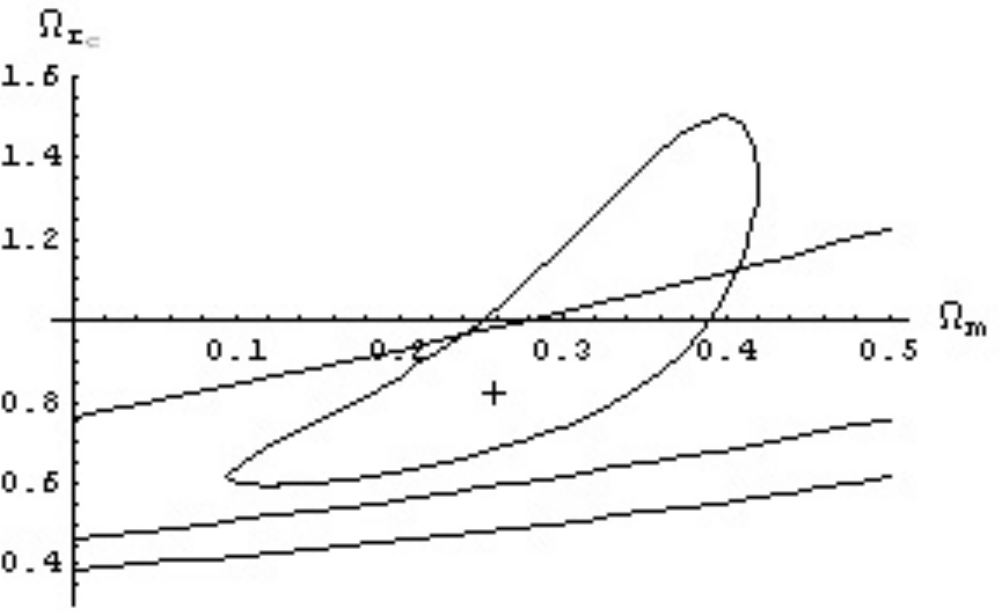}
   \caption{The closed contour is the confidence region in the $(\Omega_m^0,\Omega_{r_c})$-\newline  plane from a fit to SNeIa data. The point of best fit is marked by a\newline  cross. The curves crossing from left to right correspond to the flat-\newline space condition Eq.~(\ref{Om0}) for the upper $1\sigma$ value (top), the central val-\newline ue (middle) and the lower $1\sigma$ value (bottom) of the parameter $\Omega_A$. }
    \end{figure*}

\section{Discussion and conclusions}

To learn how well our three-parameter model fits the confidence region determined by the data, we turn to the flat-space condition, Eq.~(\ref{Om0}). Here this condition is a surface in the $\Omega_m^0,\Omega_{r_c},\Omega_A$-space which, if our model is successful, should cut the banana-shaped confidence region in Fig.1. Unfortunately the exact value of $\Omega_A$ is not known, so we must draw Eq.~(\ref{Om0}) for several values. Obviously the model is a good fit to the observational data when $\Omega_A$ is within the $1\sigma$ range quoted in Eq.~(\ref{res}).

Substituting the solution from Eq.~(\ref{res}) into Eq.~(\ref{Omrc}), the value of the cross-over scale is $r_c=0.55/\sqrt{H_0}$. The relation Eq.~(\ref{BA}) was a conjecture that could well have been different by some numerical factor. Then the fitted values of the free parameters would have changed, but a good fit could still have been obtained. Thus we do not consider that the relation Eq.~(\ref{BA}) is a fine-tuning condition. In contrast to the $\Lambda CDM$ model, to the Quintessence model, and to many other models, the present model does not imply any fine-tuning.

To compare models by using Information Criteria BIC (Schwarz \cite{Schwarz}) or AIC (Akaike \cite{Akaike}) as do Davis (\cite{Davis}) we consider extremely crude. The reason is that no information on parameter correlations is included. If one parameter pair has a correlation coefficient near 0.99, it should be counted as a single parameter.

It is easy to explain the \emph{coincidence problem} in this model. It is caused merely by the ratio of the scales of the action, the Planck scale $M_{Pl}$ on our brane and the bulk scale $M_5$. These constants just happen to have particular time-independent values which determine the DGP cross-over scale $r_c$ by the definition (\ref{rc}).

Our model should still be tested against other cosmological data, but this has to wait until CMB and BAO power spectra have been derived for five-dimensional braneworld cosmology. As for ISW data, the problems encountered in the simplest Chaplygin gas model are alleviated if not eliminated by the presence of DGP self-deceleration.

If this model meets all criteria, we have here a first indication that we live in a five-dimensional braneworld.

\emph{Note added in print.} Recently cosmologies embedding the generalized Chaplygin model in self-decelerating DGP gravity have been studied elsewhere (Bouhmadi-L\'opez \& Lazkoz \cite{Lazkoz}).
\begin{acknowledgements}
It is a pleasure to acknowledge numerous discussions with J. Nevalainen and J. Ahoranta. T. Davis has kindly let us use her SNeIa compilation, and H. Ruskeep\"a\"a has given precious help with Mathematica.
\end{acknowledgements}

 \end{document}